\documentclass[aps,prl,reprint,longbibliography]{revtex4-2}

\usepackage{amsfonts,amscd,mathrsfs,amsmath,amsthm,amssymb}
\usepackage{mathtools}
\usepackage{booktabs}
\usepackage{physics}
\usepackage{tikz}
\usetikzlibrary{quantikz2}
\usepackage{xparse}
\usepackage{graphicx}
\usepackage{float}
\usepackage{pifont}
\usepackage{enumerate}   
\usepackage{comment}
\usepackage{bm}
\usepackage{bbm}
\usepackage{caption}
\usepackage{subcaption}
\usepackage{adjustbox}
\PassOptionsToPackage{hyphens}{url}\usepackage{hyperref}
\hypersetup{colorlinks=true,linkcolor=blue!80!black,  citecolor=blue!80!black,}
\usepackage[capitalise]{cleveref}

\newcommand{\sm}{\text{-}}
\newcommand{\CNOT}{\mathrm{CNOT}}
\newcommand{\Q}{\mathcal{Q}}
\newcommand{\A}{\mathcal{A}}

\begin{document}

\title{Flexible Fault Tolerant Gate Gadgets}
\thanks{These authors contributed equally to this work.}
\author{Eric Kubischta}
\email{erickub@umd.edu} 
\author{Ian Teixeira}
\email{igt@umd.edu}

\affiliation{Joint Center for Quantum Information and Computer Science,
NIST/University of Maryland, College Park, Maryland 20742 USA}

\begin{abstract}
We design flexible fault tolerant gate gadgets that allow the data and the ancilla to be encoded using different codes. By picking a stabilizer code for the ancilla we are able to perform both Clifford and non-Clifford gates fault tolerantly on generic quantum codes, including both stabilizer and non-additive codes. This allows us to demonstrate the first universal fault tolerant gate set for non-additive codes. We consider fault tolerance both with respect to a dephasing channel and a depolarizing channel.
\end{abstract}

\maketitle

\newpage

\section{Introduction}

We present, for the first time, fault tolerant (FT) gate gadgets to achieve a universal FT gate set for non-additive (non-stabilizer) quantum error correcting codes.

In the standard setting of stabilizer codes, achieving universal FT gates usually involves two key steps: distilling a magic state and then injecting the magic state via gate teleportation \cite{teleportationGottesmanChuang,magicstatedist,Hastingsmagic,magiccost2}. However, magic state distillation protocols rely on the special relationship between the Clifford hierarchy and stabilizer codes \cite{disjointness}, so no distillation protocols for non-additive codes are known. The gate teleportation step also cannot be directly applied to non-additive codes due to the necessity of a FT measurement in the Bell basis, which typically requires a FT implementation of $\mathrm{CNOT}$. The ease of fault tolerantly implementing $\mathrm{CNOT}$ on CSS codes is a significant factor in the design of universal FT quantum computers, almost all of which rely on CSS codes. For instance, in surface codes \cite{bravyi1998quantumcodeslatticeboundary} a FT implementation of $\mathrm{CNOT}$ can be achieved via lattice surgery \cite{LatticeSurgery}. More generally, every CSS code has a transversal $\mathrm{CNOT}$; indeed, a stabilizer code has a transversal $\mathrm{CNOT}$ if and only if it is CSS \cite{gottesmanbook}. In contrast, no known non-additive code implements $\mathrm{CNOT}$ transversally.


There are some alternative proposals for universal FT gates for stabilizer codes that do not use state distillation and teleportation, but they rely on specific properties of certain stabilizer codes such as gauge subsystem codes \cite{Paetznick_2013,Anderson_2014}, concatenated codes \cite{Concatenated2014}, or pieceable fault tolerance \cite{pieceableFaultTolerance}.  There are variants of the standard gate teleportation circuit (e.g., one-bit teleportation \cite{onebitTeleportation}), but they all require a FT implementation of some entangling gate. This is necessary to entangle the ancilla block, and its encoded magic state, with the data block, and thus teleport the gate. The absence of FT entangling gates for non-additive codes precludes the application of standard state injection techniques to create FT gate gadgets for these codes.

We address this gap by demonstrating the first FT gate gadgets for non-additive codes, culminating in the first example of a universal FT gate set for a non-additive code.

Let us start by examining the following simple circuit.

\begin{figure}[htp]
    \centering
    \includegraphics[width=5cm]{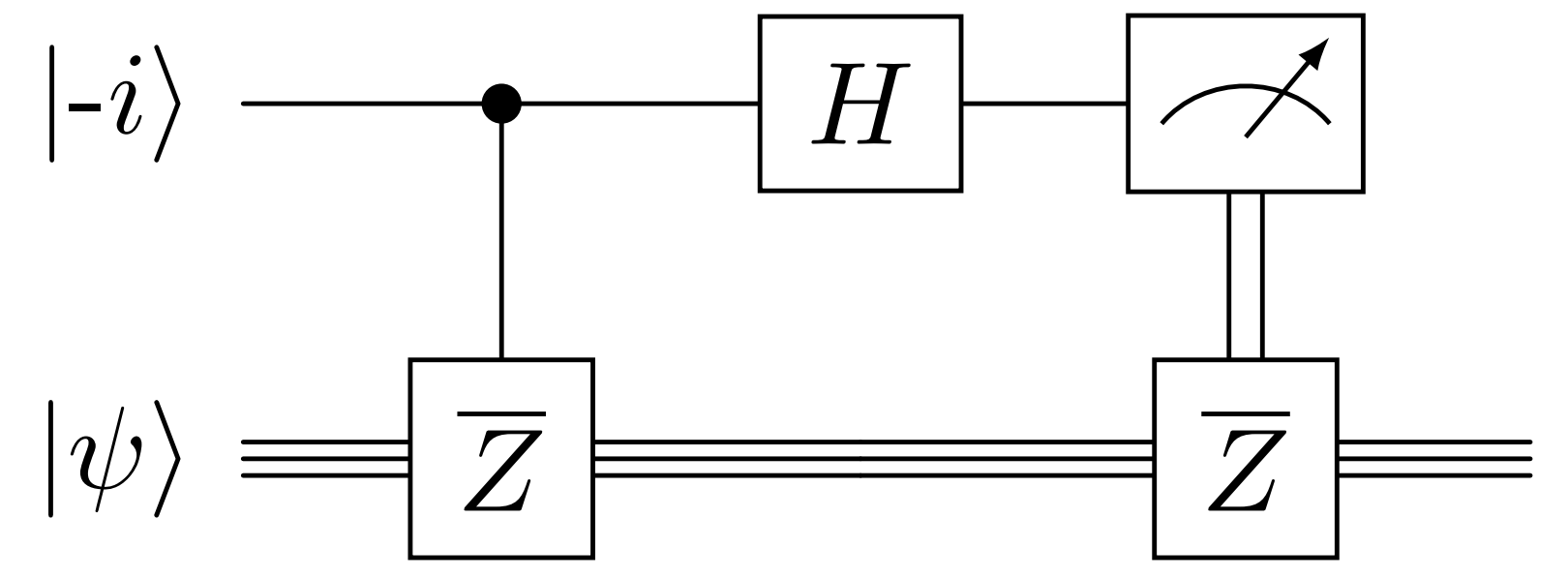}
    \caption{Non fault tolerant gate gadget implementing logical phase $\overline{S}$ for any code $\Q$ with transversal $Z$}
    \label{fig:circuit1}
\end{figure}
Here the data $\ket{\psi}$ is encoded into an $n$-qubit quantum error correcting code $\Q$ and the ancilla qubit starts in the single qubit state $\ket{\sm i} := \tfrac{1}{\sqrt{2}} \qty( \ket{0} -i \ket{1})$, which is the $\sm 1$ eigenstate of the $Y$ gate.

After the controlled logical $\overline{Z}$ gate, we are in the entangled state $\ket{0} \ket{\psi} - i \ket{1} \overline{Z} \ket{\psi}$ up to an unimportant global normalization we will ignore in the following. Then once we apply the Hadamard gate $H$ to the ancilla and rearrange we have $ \ket{0} \qty( \tfrac{I - i\overline{Z} }{\sqrt{2}} )\ket{\psi} + \ket{1} \qty(\tfrac{I + i\overline{Z}}{\sqrt{2}})\ket{\psi}$. Note that $ \tfrac{I - i\overline{Z} }{\sqrt{2}} $ is proportional to the (logical) phase gate, that is, $\tfrac{I - i\overline{Z} }{\sqrt{2}} = e^{\sm i \pi/4} \overline{S}$. Similarly, $ \tfrac{I + i\overline{Z}}{\sqrt{2}} = e^{i \pi/4}\overline{S}^\dagger$. So we can rewrite the previous as $\ket{0} \overline{S}\ket{\psi} + i\ket{1} \overline{S}^\dagger \ket{\psi}$. 

Now we measure the ancilla in the computational basis. There is a 50\% chance of measuring $ 0 $ and the system collapsing to $\overline{S}\ket{\psi} $, and a 50\% chance of measuring $ 1 $ and the system collapsing to $ \overline{S}^\dagger \ket{\psi} $. Finally, we perform a classically controlled $Z$ gate, i.e., if we measure $1$ we apply a $\overline{Z}$ gate to the data. Thus, since $Z S^\dagger = S$, the output of this circuit is (deterministically) $\overline{S} \ket{\psi}$. In other words, this circuit is a gate gadget for the logical phase gate $\overline{S}$ on the data code. 

Note that the only requirement of this circuit is that the data code $\Q$ must implement logical $\overline{Z}$ transversally as $Z^{\otimes n}$, which is a property enjoyed by almost all qubit codes, even non-additive codes. For example, this is true for many stabilizer codes and of all the non-additive codes in \cite{2004permutation,gross2,us1,us2,AydinPermutationallyInvariantQuantumCodes,us3}.

The problem with this simple circuit is that it is not fault tolerant. In practice, the controlled-$\overline{Z}$ gate would be implemented sequentially as a series of controlled-$Z_i$ gates acting on the $i$-th data bit. Thus an early $X$ or $Y$ error on the ancilla would propagate multiple $Z$ errors to the data. Meanwhile, a $Z$ ancilla error wouldn't propagate any errors to the data, but the gadget would end up implementing $\overline{S}^\dagger$ instead of $\overline{S}$. The rest of this paper is dedicated to making this simple circuit fault tolerant.

\emph{Application.--} Before we get to fault tolerance, it is natural to wonder - why is this gadget interesting at all? Isn't implementing a phase gate $S$ completely useless given that many stabilizer codes can already implement the phase gate transversally? 

The answer is that we are interested primarily in non-additive codes, which often do not have a transversal implementation of $S$, and some can be made universal upon adding such a gate. For example, the $((7,2,3))$ icosahedral code  has transversal $X$, $Y$, and $Z$, as well as $HS^\dagger$ (but neither $H$ nor $S$ individually) as well as an exotic gate (not in the Clifford hierarchy) called the $\Phi$-gate \cite{us1}. These gates together generate a \textit{maximal} finite subgroup (the binary icosahedral group) of the single qubit gates, and so adding any single gate outside this group yields a single qubit universal gate set. Since $S$ is not in the binary icosahedral group, then even a Clifford gate as simple as $S$ is enough for universality. In general, icosahedral codes that can correct a single error exist for all odd lengths $n \geq 7$ \cite{us3} and they also exist for higher orders of error correction, for example a $((31,2,5))$ icosahedral code \cite{ushigherdistance2I}. All of these icosahedral codes are necessarily non-additive \cite{us1} and in all of these codes, an $S$ gate gadget is enough for universality.

Going further, there is nothing special about $Z$ in the previous circuit (\cref{fig:circuit1}). We could just as easily use $X$ or $Y$. We do want to use a Pauli though, because nearly all known codes, both stabilizer and non-additive, have access to a transversal $X$, $Y$, and $Z$, and controlled-Pauli gates are in the Clifford group and thus conjugate Pauli errors to Pauli errors which simplifies the error analysis. Using a $\overline{Y}$ in the previous circuit instead of a $\overline{Z}$ produces the gate $\tfrac{1}{\sqrt{2}}(I - i \overline{Y})$ which is proportional to $\overline{X} \overline{H}$. So post-multiplying the data by a transversal $\overline{X}$ would give us a gate gadget for the logical Hadamard gate $\overline{H}$. Note that $ S $ and $ H $ together generate the single qubit Clifford group, so this is sufficient to allow FT implementation of local Clifford gates on any (even non-additive) code with transversal Pauli gates.

Even some stabilizer codes can be made universal with this $H$ gadget, for example the $[[15,1,3]]$ code has $X$, $Z$, $S$, $T$, and $\CNOT$ transversal and so adding in an $H$ gate gadget would yield universality. In fact, for all the CSS codes with transversal $ T $ (the so called triorthogonal codes \cite{smallestT,smallestT2,smallestT3}) an $H$ gate is sufficient for universality \cite{Paetznick_2013}.

Lastly, we could even change the ancilla input state. Suppose instead of $\ket{\sm i}$ we prepared the ancilla in the state $\ket{\tfrac{\pi}{8}} := \cos(\tfrac{\pi}{8}) \ket{0} -i \sin(\tfrac{\pi}{8}) \ket{1}$. Then we would implement the gate $\cos(\tfrac{\pi}{8}) I - i \sin(\tfrac{\pi}{8}) \overline{Z}$ which is proportional to the non-Clifford gate $\overline{T}$. The $\ket{\tfrac{\pi}{8}}$ state might seem obscure but it is simply the $+1$ eigenvector of the Clifford gate $S^\dagger H S$.  

In summary, small variations of this simple circuit allow one to implement a variety of gates on \textit{any} code that supports a transversal implementation of the Pauli group (including non-additive codes). In what remains, we will show various ways and scenarios in which these circuits can be made fault tolerant.

\section{Fault Tolerant Gate Gadget for Dephasing Noise}

One of the reasons the circuit in \cref{fig:circuit1} isn't fault tolerant is that the ancilla is a single qubit and so a single ancilla error can corrupt the gadget's output. One way to rectify this is to encode the ancilla into a quantum error correcting code. Consider then the following circuit.

\begin{figure}[htp]
    \centering
    \includegraphics[width=\linewidth]{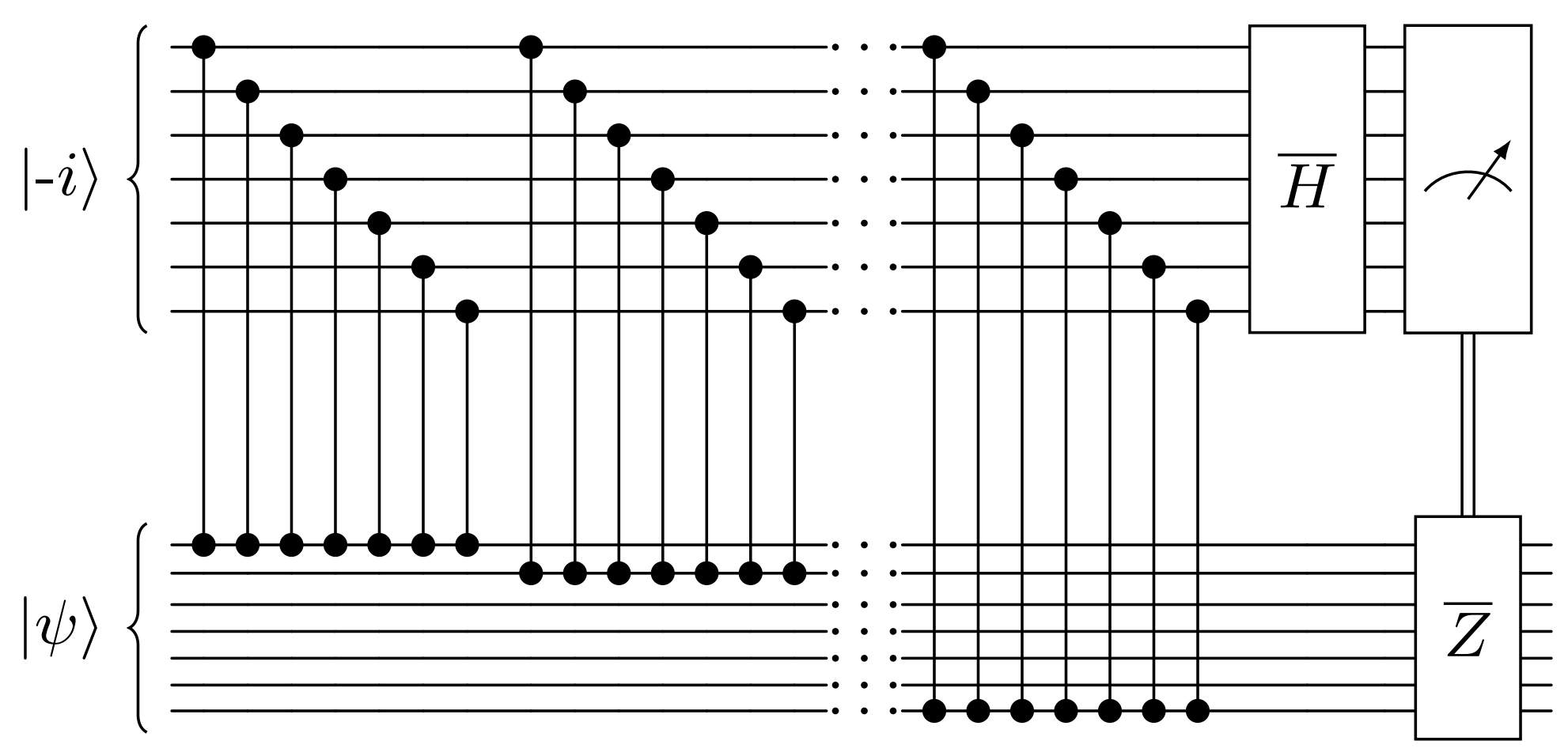}
    \caption{FT gate gadget (for $Z$-biased noise) implementing logical phase $\overline{S}$ for any code with transversal $\overline{Z}$}
    \label{fig:circuit2}
\end{figure}

Here $\ket{\psi}$ is the data encoded into a distance $d$ qubit code $\Q$ of length $n$ (perhaps non-additive) that implements logical $\overline{Z}$ transversally as $Z^{\otimes n}$. The ancilla is prepared in the logical $\ket{\sm i}$ state of any distance $d$ stabilizer code $\A$ of length $m$. Note that the ancilla code $\A$ and the data code $\Q$ do not have to be the same, nor do they even have to be the same length (in stark contrast to gate teleportation). In \cref{fig:circuit2} we use two connected dots to represent a controlled-$Z$ gate. So we perform a controlled-$Z$ from each of the $m$ ancilla qubits to the 1st data qubit in series, call this round 1, then we do the same thing for the 2nd qubit of the data, call this round 2, etc. So there are $n$ total rounds with $m$ controlled-$Z$'s in each round giving a total of $nm$ controlled-$Z$'s. Then we perform a FT Hadamard on the ancilla code, ideally using a transversal implementation. For example, in any CSS$(C_X,C_Z)$ code with $C_X = C_Z$, it is the case that $H^{\otimes m}$ implements logical $\overline{H}$ because the physical gate just swaps all of the $X$-type stabilizers with the $Z$-type stabilizers. Lastly we use a FT sub-gadget to correct the existing ancilla errors and measure $\overline{Z}$, for example using Shor error correction \cite{ShorEC}, Steane error correction \cite{SteaneEC}, or Knill error correction \cite{KnillEC}. And if we measure $1$ we perform a transversal $\overline{Z}$ on the data.

The key insight as to why this circuit works is that any code with $Z$ exactly transversal has a special parity property. In particular, logical $\ket{0}$ is a superposition of even-weight bit strings and logical $\ket{1}$ is a superposition of odd-weight bit strings. So during the $i$-th round, logical $\ket{0}$ turns on the controlled-$Z$ for the $i$-th data qubit an even number of times (effectively not implementing a gate) and logical $\ket{1}$ turns on the controlled-$Z$ on the $i$-th data qubit an odd number of times (effectively implementing a single $Z$). So after all $n$ rounds, we obtain the state $\ket{0} \ket{\psi} - i \ket{1} \overline{Z} \ket{\psi}$, the same as in \cref{fig:circuit1}. If we black box all of the controlled-$Z$ gates we have effectively constructed a logical controlled-$\overline{Z}$ sub-gadget that acts between two different codes. Thus the rest of \cref{fig:circuit2} proceeds as in \cref{fig:circuit1} and we have verified that \cref{fig:circuit2} is indeed an $\overline{S}$ gate gadget.

A dramatic improvement in the circuit complexity of \cref{fig:circuit2} can be achieved by only hooking up those qubits that support the smallest $Z$-type logical operator. For example, in the $[[7,1,3]]$ Steane code \cite{SteaneCode}, $IIIZZZZ$ is a stabilizer and using the fact that $Z^{\otimes 7}$ implements logical-$\overline{Z}$ we have that $ZZZIIII$ also implements logical-$\overline{Z}$ (and this is the lowest weight logical operation possible since the code has distance $3$). This means the first 3 qubits of the Steane code also have a special parity property, i.e., logical $\ket{0}$ is a superposition of kets such that the first 3 bits are even and logical $\ket{1}$ is a superposition of kets such that the first 3 bits are odd. So each round is only composed of $3$ controlled-$Z$ gates instead of $m=7$. So in aggregate we only need $3n$ controlled-$Z$ gates instead of $mn$. In fact, if a code has distance $d$ then the smallest weight a logical operator can be is $d$, so a useful and easily attainable lower bound on the number of controlled-$Z$ gates is $dn$.

\emph{Evaluating Fault Tolerance.--} A quantum channel for which either no error happens or a fully Z-biased error happens is called a quantum dephasing channel \cite{Watrous}. Dephasing is the dominant source of noise in many hardware systems used for quantum computation, including superconducting qubits, quantum dots, and trapped ions \cite{Dephasing1,Dephasing2,Dephasing3,Dephasing4}. This bias towards dephasing noise is even present in bosonic systems \cite{DephasingBosonic}. This has motivated the design of many FT constructions for $ Z $-biased noise, for example tailoring the stabilizers for surface codes to protect against $Z$-biased noise \cite{DephasingUltrahigh}, and tailoring decoders for $Z$-biased noise \cite{DephasingDecoder}. 

\cref{fig:circuit2}, like \cref{fig:circuit1}, still suffers from early $X$ or $Y$ ancilla errors propagating many $Z$ errors to the data. However, it is now fault tolerant against $Z$ errors. A $Z$ error on either the data or ancilla will not propagate any errors, since $Z$ errors commute with controlled-$Z$ gates. Moreover, the measurement subgadget will catch and correct ancilla errors so as not to corrupt the output of $\overline{S}$ (this was our original motivation for using an error correcting code as the ancilla). Thus \cref{fig:circuit2} is a FT gate gadget for dephasing channels.

The logical $\ket{\sm i} $ state can be fault tolerantly prepared using standard methods since it is the $\sm 1$ eigenstate of logical $\overline{Y}$ for a stabilizer code \cite{gottesmanbook}. Alternatively, if the ancilla code $\A$ implements $S$ transversally, we can prepare the logical $\ket{0}$ state fault tolerantly, and then transversally apply $\overline{S}^\dagger \overline{H}$ since $\overline{S}^\dagger \overline{H} \ket{0} = \ket{\sm i}$.

Also note that by changing the input ancilla state to the logical $\ket{\tfrac{\pi}{8}} = \cos(\tfrac{\pi}{8}) \ket{0} -i \sin(\tfrac{\pi}{8})\ket{1}$ state, we can implement the logical $\overline{T}$ gate on the data. Such a state can be prepared fault tolerantly in a similar fashion to stabilizer states since logical $\ket{\tfrac{\pi}{8}}$ is the $+1$ eigenstate of the Clifford gate $\overline{S H S^\dagger}$ \cite{gottesmanbook}. Also magic state distillation techniques are available because the ancilla code $\A$ is a stabilizer code.

\section{Fault tolerant gate gadget for Depolarizing Noise}

\begin{figure}[htp]
    \centering
\includegraphics[width=\linewidth]{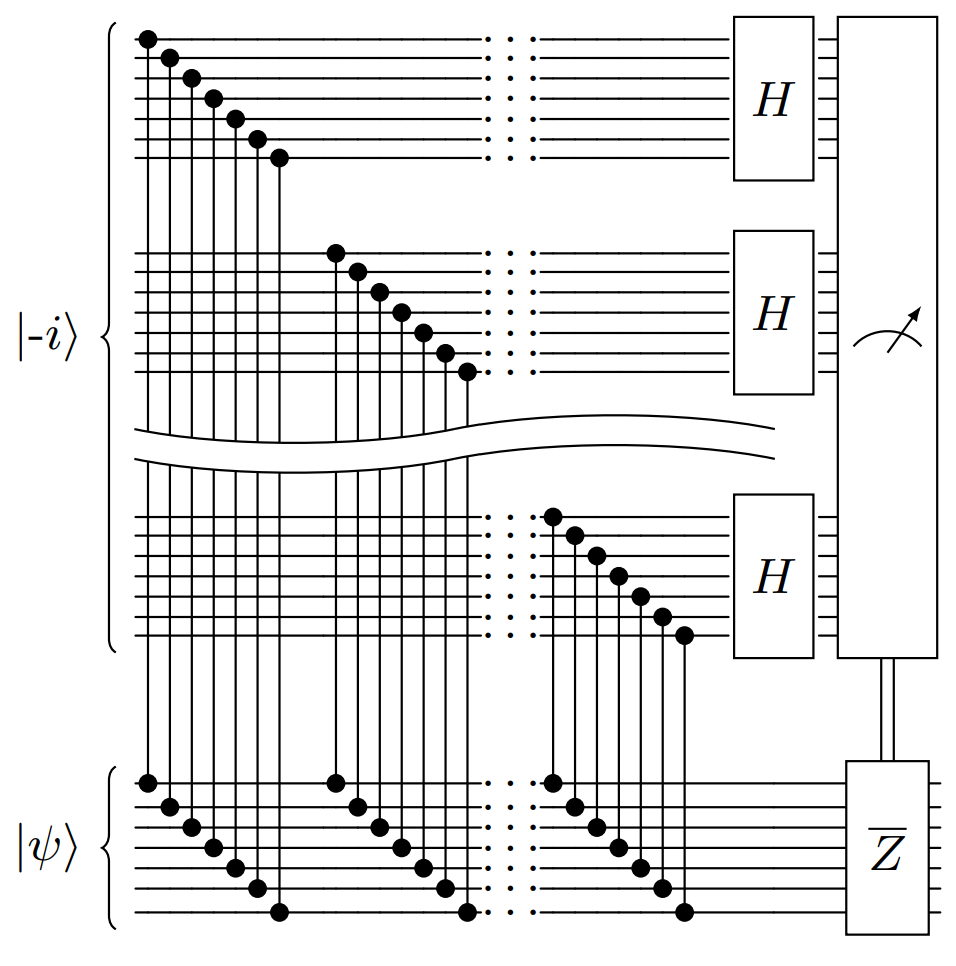}
    \caption{Fault tolerant gate gadget implementing logical phase $\overline{S}$ for any code with transversal $Z$}
    \label{fig:circuit3}
\end{figure}

One critique of \cref{fig:circuit2} is that the ancilla block is not hooked up to the data block transversally, so $X$ and $Y$ errors will inevitably propagate $Z$ errors to many data legs. This can be rectified by the circuit in \cref{fig:circuit3}.

Again $\ket{\psi}$ is the data encoded into an $n$-qubit code $\Q$ of distance $d$ that implements $\overline{Z}$ transversally. Let $\A_1$ be any $m$-qubit code of distance $d$ that supports transversal $Z$ and $H$ as in \cref{fig:circuit2}. Let $\A_2$ be an $n$-qubit repetition code. Then we take the ancilla code $\A$ in the above circuit to be a concatenation between $\A_1$ and $\A_2$, that is, $\A = \A_2^{\otimes m} \circ \A_1$ noting that $\A$ has length $mn$. We prepare the ancilla code $\A$ into the logical $\ket{\sm i}$ state.

Using a concatenated code allows us to still exploit the special parity property of the circuit in \cref{fig:circuit2} whilst hooking up the circuit in a transversal manner. Namely, the $i$-th repetition code $\A_2$ is hooked up transversally to the data during the $i$-th time step. So after all $m$ time steps, we are in the state $\ket{0} \ket{\psi} - i \ket{1} \overline{Z} \ket{\psi}$ just as in the previous two circuits. Then we implement a transversal Hadamard $\overline{H}$ for the $\A_1$ code by implementing (a possibly non-fault tolerant) Hadamard $H$ on each of the $m$ copies of the $\A_2$ repetition codes; this overall process is still fault tolerant as pointed out in \cite{Concatenated2014}. At this point we are in the same position as the previous two circuits and so we have another $\overline{S}$ gate gadget.

We can use the reduced parity principle as in \cref{fig:circuit2} to shrink the number of time steps from $m$ to as few as $d$ time steps (if the $\A_1$ code supports an implementation of $\overline{Z}$ using a $Z$-type operator of weight $d$ like the $[[7,1,3]]$ code). Notice that \cref{fig:circuit3} is much shorter in duration (at the cost of overhead) than \cref{fig:circuit2}; it only requires $m$ time steps for the controlled-$Z$ gates (and $d$ time steps in the best case) as opposed to $mn$ time steps in \cref{fig:circuit2} (or $dn$ time steps in the best case). So there is an $n$-fold improvement in duration.

\emph{Evaluating Fault Tolerance.--} In a similar way to \cref{fig:circuit2}, the circuit in \cref{fig:circuit3} is fault tolerant for fully $Z$-biased noise. We also have additional improvement because the ancilla blocks are hooked up transversally to the data block. Namely, we are now fault tolerant for any type of noise on the ancillas. This allows us to consider a more standard error model, the depolarizing channel, where $X$, $Y$, and $Z$ errors are equally likely.

 At first glance, \cref{fig:circuit3} is not fault tolerant for all types of noise on the data. Suppose an $X$ or $Y$ error happens on the 1st data qubit at the start of the circuit. It will propagate a $Z$ error to the 1st qubit of each of the $m$ repetition codes which will end up causing a logical $\overline{Z}$ error on the full ancilla code $\A$, corrupting the output of the gadget by implementing $\overline{S}^\dagger$ rather than $\overline{S}$.

One solution is to interleave FT error correction sub-gadgets on the data code $\Q$ between each of the controlled-$Z$ rounds. In this manner, one could prevent a $Z$ error from being propagated to many of the repetition codes and thus prevent a logical error being built up for the full ancilla code $\A$.  For example, after the 1st round of controlled-$ Z $ gates we have the product state $ \ket{v_0} \ket{\psi} - i \ket{v_1} \overline{Z} \ket{\psi}$ where $\ket{v_0}$ is the state of $\A_1$ that consists of all kets that have $0$ as the 1st qubit and $\ket{v_1}$ consists of all kets that have $1$ as the 1st qubit. However, because $\ket{\psi}$ and $\overline{Z} \ket{\psi}$ are within the same codespace (since $\overline{Z}$ preserves the codespace), measuring and correcting the data errors won't destroy the superposition. The same is true after each round. The upshot of this is that if you interleave FT error-correction or FT measurement gadgets between each data round, this circuit is a fully FT gadget for any data code that admits a transversal implementation of $\overline{Z}$. Moreover, one could replace the controlled-$Z$ gates with controlled-$X$ or controlled-$Y$ gates (as mentioned in the introduction) to construct other fully FT gate gadgets (such as an $H$ gadget).

For stabilizer codes, there are an abundance of FT error correction and measurement gadgets. However, the situation for non-additive codes is more complex. To date, there has been minimal research on the fault tolerance of non-additive codes, and no definitive FT gadget for error correction exists. Nonetheless, the method presented in \cite{ouyangFTmeas} offers a robust approach for measuring non-additive errors, though its complete fault tolerance remains to be fully evaluated. To achieve fully FT non-additive codes, a FT error correction gadget is essential. The development of such a gadget, when it occurs, will be instrumental and can be incorporated as a sub-gadget here.

\section{Conclusion}

We have introduced innovative FT gate gadgets that can be applied to any quantum code supporting a transversal implementation of the Pauli gates, including non-additive codes. Our circuits provide an alternative to gate teleportation by allowing for the ancilla code to differ from the data code, even in length, potentially reducing overhead in the high-distance limit by minimizing the size of the ancilla code. A significant insight is that for codes where $Z$ is exactly transversal, a special parity property enables the implementation of a logical controlled-$Z$ gate (or controlled-$X$ or -$Y$) across two different codes. The versatility of our circuits stems from the fact that nearly all quantum codes support a transversal implementation of the Pauli group. We are optimistic that our work will spur further research into fault tolerance for non-additive codes, including FT encoding, decoding, state preparation, measurement, and error correction.

\section{Acknowledgements}
We thank Daniel Gottesman for helpful conversations regarding fault tolerance and for early access to his textbook \cite{gottesmanbook}. We thank Michael Gullans for helpful conversations and for encouraging us to look into fault tolerance for non-additive codes. We also thank the members of Quantum Computing Stack exchange for helpful insights regarding various aspects of fault tolerance. We used Quantikz \cite{quantikz} to prepare the circuits in this paper. This research was supported in part by the MathQuantum RTG through the NSF RTG Grant No. DMS-2231533.

\bibliography{biblio} 

\end{document}